\documentstyle[manuscript,aps]{revtex}
\begin{document}
\draft
\title{On Hypothesis of the Two Large Extradimensions}
\author{Janusz Garecki}
\address{Institute of Physics, University of Szczecin, Wielkopolska 15; 70-451
Szczecin, POLAND}
\date{\today}
\maketitle
\begin{abstract}
Recently there was proposed a hypothesis about existence of the two large
extradimensions. This hypothesis demands, e.g., modification of Newton law at
submilimeter scale. In this brief report we  show that this hypothesis cannot be
correct in present formulation.
\end{abstract}

\section{The hypothesis, its consequences and conclusion}
In order to solve the so-called ``hierarchy problem'' and give an
explanation why gravity is so weak in comparison with other known
interactions some authors [1--5] proposed recently to use {\it large
extradimensions}. Namely, these authors assert that the gravity is
really strong and the {\it electroweak unification energy scale} $M_u =
1$ Tev ($ = 10^{19} m^{-1}$ in natural units in which $\hbar = c = 1$)
and the fundamental Planck's scale $M_P = 10^{16}$ Tev ($= 10^{35}
m^{-1}$) {\it are indeed the same size}, but four-dimensional gravity is
so weak (hence $M_P$ is so large) due to dilution of gravity (Why only
gravity?) in large extradimensions. So, these authors assert that the
unification energy $M_u = 1$ Tev {\it is the only fundamental scale} in
Nature. If so, new dimensions, black holes, quantum gravity, and string
theory will become  experimentally accessible in near future [6--8].

Following [1--4] we have the following formula on radii $R_c$ of
compactified extradimensions
\begin{equation}
R_c = {\hbar\times c\over M_u}\biggl({M_P\over M_u}\biggr)^{{2\over n}},
\end{equation}
where $n$ denotes number of compactified extradimensions. ${M_P\over
M_u} = 10^{16}$ if $M_u = 1$ Tev.

From (1) one can easily get that if:
\begin{enumerate}
\item $n = 1$, then $R_c \approx 1 AU$ ,
\item $n=2$, then $R_c = 10^{-3} m$,
\item $n = 7$, then $R_c =10^{-14}m$.
\end{enumerate}

The most popular is the second case in which two spatial extradimensions
are curled up into circles about $10^{-3}$ m in size. In this case we have
five-dimensional space and six-dimensional spacetime. 

The first possibility must be rejected at once because it is from the
beginning {\it incompatible with experience}. 

As we will show the second possibility {\it also should be rejected} at
least from the two following causes.

Firstly, it was shown in past be P. Ehrenfest,
G.J. Whitrow and others [9--11] that the three-dimensional space, i.e.,
four-dimensional spacetime {\it is necessary for many reasons}, e.g.,
only in a three-dimensional space atoms can be stable. So, if hypothesis
about two large extradimensions which have radii $R_c = 10^{-3} m \gg
10^{-10}m$ ($10^{-10}m$ is a typical diameter of an atom) were correct
than {\it our existence would be impossible}. 

Secondly, if the hypothesis on two large spatial extradimensions was
correct then the Newton law {\it had to be changed} in the scale $r\leq
R_c = 10^{-3} $ m. It can be most easily seen by use of the {\it Gauss
law} in $N-$ dimensional space for a point mass $m$
\begin{equation}
\int\limits_{x_1^2 + x_2^2 + ... + x_n^2 = R^2} \vec{E}\cdot\vec{n}
d\sigma = m.
\end{equation}
Here $\vec{E}$  means the gravitational strength, $\vec{n}$ is the unit
normal to the imagined Gauss sphere and $d\sigma$ denotes an integration
element over this sphere.

Using spherical symmetry one can easily obtain from (2)
\begin{equation}
E(R)\int\limits_{x_1^2+x_2^2+ ... + x_n^2} d\sigma = E(R)
{N\pi^{N/2}R^{N-1}\over\Gamma(N/2 +1)} = m.
\end{equation} 

From (3) there follow
\begin{equation}
E(r) = {G_N m\over r^{N-1}},
\end{equation}
and {\it the modified Newton law} for the value of the gravitational
force $F$ between two point masses $m_1, ~m_2$ if $r\leq R_c = 10^{-3}$ m
\begin{equation}
F = {G_N m_1 m_2\over r^{N-1}}.
\end{equation}
Here $G_N$ denotes a new gravitational constant.
We have [1--4] 
\begin{equation}
G_N = \bigl(M_u\bigr)^{-(2+n)}, ~~M_u = 1 Tev (= 10^{19} m^{-1}),
\end{equation}
where $n = N-3$ is the number of the curled up spatial extradimensions.

For $N=5$ ,i.e., for $n=2$,we get from (5-6)
\begin{equation}
F = {G_N m_1 m_2\over r^4},~~G_N = (M_u)^{-4}= 10^{-76} m^4.
\end{equation}
In the following we will confine ourselves to the last, most popular
possibility when $n = 2$, i.e., we confine to the five--dimensional space
and to the six--dimensional spacetime.

We will show  that the modification Newton law for $r\leq R_c = 10^{-3}$
m given in the case by formula (7) {\it cannot be correct}.

With this aim we will use an old Stanford experiment [12--13] on free
falling conductivity electrons inside of a freely standing or freely
hanging metal ( inside Cu).
This experiment, performed with very high precise, showed that the
conductivity electrons in such a metal (Cu) were falling
(under influence of the Earth gravitational field) {\it with the
same acceleration } $g_{int}$ {\it inside metal} (Cu) {\it as in vacuum},
i.e., they showed that $g_{int} = g_{ext} = 9,8 {m\over s^2}$.

Such result is okay if we apply {\it the same, ordinary} Newton law to
gravitational interaction between an electron inside of the metal and Earth
and to gravitational interaction of this electron and a positive ion of the
cristal lattice of the metal (Cu). Namely, the simple calculation shows
that for {\it ordinary Newton law} the ratio of the values of the
gravitational forces between electron-Earth ($F_{e-E}$)
and between  electron-ion ($F_{e-ion}$) is equal 
\begin{equation}
{F_{e-E}\over F_{e-ion}}\approx 0.3\times 10^{16}.
\end{equation}
Calculating the ratio (8) we have taken $ m_{ion} = 108\times 10^{-27}
kg, ~M_{Earth}=: M_{E} = 6\times 10^{24} kg,~ R_{Earth}=: R_{E} =
6,4\times 10^{6} m$ ($R_{E}$ = distance between an electron inside of the
metal and the center of the Earth)  and $r = 0,5\times 10^{-10}$m  as an
upper value of the distance between a conductivity electron and the 
nearest positive ion inside of the metal (of course the distance between
conductivity electrons and positive ions inside of the metal can be smaller
than the last value).

The result (8) shows that the ordinary Newton gravitational interaction
between conductivity electrons and positive ions of the metal {\it is
neglegible} in comparison with ordinary Newton's gravitational
interaction between the same electrons and Earth. In consequence, the
conductivity electrons inside of a metal can freely fall with
the same acceleration as in vacuum, and give an uniform, gravity induced
electrostatic field inside of the metal [12,14]. This gravity induced
electrostatic field is in an equilibrium with the gravitational field
which acts between an electron inside a metal and Earth.

However, if the hypothesis about the two large extradimensions curled up
to the size $R_c = 10^{-3}$ m is correct, then {\it one should change Newton
law} for gravitational interaction between a conductivity electron and a
positive ion of the metal cristal lattice to the form (7), i.e., to the form
\begin{equation}
F_{e-ion} = {G_N m_e\times m_{ion}\over r^4}, ~~G_{N}  = (M_u)^{-4} = 10^{-76}
m^4, 
\end{equation}
because an upper limit of the distance distance, $r$,  between interacting particles is in
the case $r\approx 0,5\times 10^{-10} m\ll 10^{-3}m = R_c$ (and, of course,
the distances between conductivity electrons and positive ions in
general can be smaller). On the other hand, the Newton law for 
gravitational interaction between a conductivity electron and Earth {\it
should be unchanged}, i.e., it should have the ordinary form
\begin{equation}
F_{e-E} = {G m_e\times M_E\over {R_E}^2}, ~~G = (M_{Pl})^{-2} =
10^{-70}m^2, 
\end{equation}
because in this case the distance between interacting bodies
(electron--Earth) is of order $R_E = 6,4\times 10^6$ m $\gg R_c =
10^{-3}$m. 

After doing so one can easily calculate that then, the ratio of the
values of the gravitational forces 
\begin{equation}
{F_{e-ion}\over F_{e-E}}
\end{equation}
is greater than $1$ already for $r\leq 3\times 10^{-11}$ m, e.g., if
$r = 3\times 10^{-11}$ m, then this ratio is $\approx 74> 1$. 
Here $r$ means a distance between an conductivity electron and an ion of
the cristal lattice of the metal.

But in such situation the conductivity electrons {\it could not freely
fall} inside a metal in the same way as they do in vacuum,
i.e., with the same uniform acceleration as they freely fall in vacuum. The
conductivity electrons inside of a metal {\it should fall with an
effective } acceleration  $g_{eff} = g_{int}\not= g_{ext} = 9,8{m\over
s^2}$ and they rather should fall ``onto'' positive ions of the cristal
lattice instead of onto Earth. Especially in a near vicinity of the metal walls and when
if there would be cristal lattice defects inside of the metal.

Thus, we conclude from this  old Stanford experiment that {\it one cannot change Newton law at
least up to distances} $r\approx 10^{-11}$ m in the manner expected  by the
hypothesis about two large extradimensions\footnote {Of course there are
always possible modifications which follow from general relativity}. The radii $R_c$ of the curled up into
circles two spatial extradimensions, if they really exist, {\it should be smaller}
than $R_c = 10^{-11}$m , i.e., they should be much smaller than recently
proposed $R_c = 10^{-3}$m [1--8], and, in consequence, the unification energy
scale {\it should be much greater} than $M_u = 1$ Tev.

It is interesting in this context that the recent experiments [14,15]
discovered {\it no deviations} from Newton's law up to distances
$r\approx 0,2 mm$. 

Of course, one can preserve $M_u = 1$ Tev as an admissible unification
energy scale but this demand {\it increasing} of the number of the
compactified extradimensions to seven,i.e., this demand
eleven-dimensional spacetime. Then the  radii $R_c$ of these seven 
extradimensions should be curled up, as we have mentioned about that already, to the
size $R_c\approx 10^{-14} $m.

Summing up, we think that the hypothesis about two large spatial
extradimensions with $R_c\approx 10^{-3}$ m and about $M_u = 1$ Tev as
the grand unification energy scale {\it is incorrect}.
Our existence and experiments {\it contradict this hypothesis}.

If the extradimensions really exist (this is very problematic and
controversial, see eg., [16]), then they should be compactified in a much 
smaller scale than the proposed scale $R_c \approx 10^{-3}$ m, e.g., in the
scale $R_c\leq 10^{-14}$m. The most probably they should be compactified in the Planck's
scale, i.e., in the scale originally proposed. But then we return back to
the ``hierarchy problem'' and to the cosmological constant problem [16].

\end{document}